\documentclass[pre,reprint,showkeys]{revtex4-2}

\usepackage{mathtools}
\usepackage{graphicx}

\begin{document}

\title{ Radicalism:\\ \hspace*{-0.8cm}\makebox{The asymmetric stances of Radicals versus Conventionals} }

\author{Serge Galam}
\email{serge.galam@sciencespo.fr}
\affiliation{CEVIPOF - Centre for Political Research, Sciences Po and CNRS, 98 rue de l'Universit\'e Paris, 75007, France}
\author{Richard R.W. Brooks}
\email{richard.brooks@nyu.edu}
\affiliation{Emilie M. Bullowa Professor of Law, Law School, New York University, New York, USA}

\date{March 08, 2022}

\begin{abstract}

We study the conditions of propagation of an initial emergent practice qualified as extremist within a population adept at a practice perceived as moderate, whether political, societal or religious. The extremist practice is carried by an initially ultra-minority of Radicals (R) dispersed among Conventionals (C) who are the overwhelming majority in the community. Both R and C are followers, that is, agents who, while having arguments to legitimize their current practice, are likely to switch to the other practice if given more arguments during a debate. The issue being controversial, most C tend to avoid social confrontation with R about it. They maintain a neutral indifference assuming it is none of their business. On the contrary, R aim to convince C through an expansion strategy to spread their practice as part of a collective agenda. However, aware of being followers, they implement an appropriate strategy to maximize their expansion and determine when to force a debate with C. The effect of this asymmetry between initiating or avoiding an update debate among followers is calculated using a weighted version of the Galam model of opinion dynamics. An underlying complex landscape is obtained as a function of the respective probabilities to engage in a local discussion by R and C. It discloses zones where R inexorably expand and zones where they get extinct. The results highlight the instrumental character of above asymmetry in providing a decisive advantage to R against C. It also points to a barrier in R initial support to reach the extension zone. In parallel, the landscape  reveals a path for C to counter R expansion pushing them back into their extinction zone. It relies on the asymmetry of C being initially a large majority which puts the required involvement of C at a rather low level.
\end{abstract}


\maketitle

\section{Introduction}

In recent decades, radicalism has been prevalent all over the world. In particular, radicalism has fueled a series of violent activities in many countries, prompting governments and public agencies to allocate considerable resources to implement counter-radicalization policies \cite{counter1, counter2}. These efforts, however, have so far shown limited results  \cite{fail} and despite an abundant corpus of research on the subject, the mechanisms of radicalization remain poorly understood. Indeed, although radicalization is an old phenomenon, there is still no single, recognized definition, because in cases where it occurs its qualification varies depending on the parties involved in the conflict \cite{wiki}. 

In this paper, we address the issue of radicalism by modeling the phenomenon within a simplified description of society. The purpose of our modeling is to shed light on certain aspects of the phenomenon being aware that our ideal description, as well as the associated results, should not be taken as absolute, but as indicators of trends that might be at work in the real world. 

Accordingly, we consider a simplified society governed by a set of rules, norms and cultural behaviors, which are largely followed and respected by most of its members, labeled conventionalists (C) in our model. However, it may happen that a tiny minority of members of this society start to oppose some of the established rules advocating alternative ones, which conventionalists see as extremist as opposed to the current ones perceived as moderate.  C label those people as radicals (R).  Often, R frame their challenge within broader ideals, which may be political, social or religious.

More precisely, we study the conditions under which an extremist view advocated by a tiny initial minority of R can spread through the society by persuading conventionalists to abandon the established view and become themselves radicals. The driving dynamic by which C become R is what we define as radicalization. We also explore the dynamics whereby members of the minority of R are themselves persuaded to abandon their extremist views to become C again, that is, counter-radicalization.

Our model assumes that C and R are not rigid in the view they hold, but are ``followers", i.e., agents who are subject to being persuaded to change their stance. In particular, when discussing the issue of contention in small groups of people, followers may switch their current stance to the opposite one when given more convincing arguments than those they use to legitimize their current stance. This shifting feature is implemented using the Galam model of opinion dynamics, which updates individual visions in discussing groups by applying a local majority rule with every agent holding one argument, all arguments having the same weight to convince \cite{mino}. 

However, R and C do not share the same propensity to engage in conversation over the contested practice or belief. On the one hand, most C are reluctant to provoke a debate on an issue they perceived as controversial. They maintain a neutral indifference about the subject assuming it is none of their business. On the other hand, R are eager to persuade C by debating the merits of the competing views with an aim toward expansion of their vision. Indeed, they plan to disseminate their vision as part of a collective program. Yet, cognizant of their own susceptibility to persuasion, since they are ``followers" like C, they tend to avoid group discussions where they are outnumbered. That is a key difference among agents in our model with C rarely instigating a debate while R do often instigate debate when they are the majority in the group but not when in the minority. Nevertheless, both C and R agree to discuss the issue when raised by one member of the group.
 
To account for the effect of above asymmetric propensity to engage in a debate within a group of people, we expand the Galam Model of opinion dynamics by weighting the local majority rules, which yields the Weighted Galam Model (WGM) \cite{voting,  chopa, mino, taksu, toth}. The WGM is part of sociophysics \cite{ brazil, frank, book, santo, bikas}, which includes numerous works devoted to the dynamics of opinion \cite{2-3, nuno1, sen, malarz, andre, nuno2, red, kasia1, bellomo, bolek, cheon, celia, bagnoli, carbone, kasia2, zanette, marco, andre-serge, iglesias, mobilla, fasano}. 

The WGM reveals a complex non linear landscape for the dynamics of opinion driven by asymmetric tipping points for the competing  extremist and moderate versions of the practice or belief at stake. R are found to spread inexorably in specific areas of the parameter space but get extinct in others. in particular, given tiny initial proportions of R, the associated critical values for the respective probabilities of initiating a local discussion by R, are identified to shift the dynamics from an area of extinction to an area of expansion

Nevertheless, this scenario is not always feasible. While a rather high degree of R involvement can compensate their small number of followers, that is not always sufficient to spread, since even with a high proportion of neutrality from C, R need to reach a certain threshold of initial support above the associated tipping point to avoid shrinking through the discussions. In addition, when R get beyond their tipping point and spread, the process of turning C to R is at first very slow during several updates of the dynamics, before it becomes fast at once. This feature explains why when the occurrence of R  becomes visible, it is often perceived as sudden whereas a long process took place beforehand.

Conversely, the topology of the dynamic landscape also reveals ways to counter radicalization. In particular, it shows that for C, having an overwhelming global numerical superiority, the required level of individual involvement to confront the issue and eventually push R into an extinction zone is low.

Our results highlight the instrumental role of two asymmetries in the faith of radicalization, which are the asymmetry in involvement versus the asymmetry in proportions. While R are aware and take benefit from the first one to compensate for their numerical weakness, C must become aware of the second one to reallize that they could stop the spread of R with a much liower involvement than the one deployed by R. Yet, without this minimum involvement from C to initiate a debate about the issue when being majority in gathering in private meetings, curbing radicalism is doomed to fail.

This conclusion extends and complete the findings of a previous work where the population was divided between a core of inflexibles and a sensitive part which could be divided between peaceful agents and opponents \cite{radi}. There, the involvement of core agents was found to be instrumental to curb the radicalization within the sensitive population. 

It is worth noting that other strategies involving geometrical correlations among a tiny proportions of agents have been investigated to explain the spreading of rumors and securing the top position in bottom-up hierarchies \cite{corre1, corre2}.

The rest of the paper is organized as follows:  the update equation of the Weighted Galam Model of opinion dynamics is derived in Section 2. A tipping point dynamics is revealed, allowing us to build the landscape of the dynamics. Section 3 investigates the conditions required for R to spread by convincing C.  The key for C to counter R expansion is elaborated in Section 4 and Section 5 contains concluding remarks.

\section{Setting up the WGM update equation}

We consider a heterogeneous community composed of Radicals (R) and Conventionals (C) with at time $ t $ the respective proportions $p_t$ and $(1-p_t)$. Each agent holds one of two competing practices, within a framework which can be political, social or religious. The stance advocated by R is qualified as extremist as opposed to the position defended by C qualified as moderate. 

The associated dynamics between R and C  is monitored using the Galam model of  opinion dynamics \cite{taksu}. In the basic Galam model agents discuss the issue at stake via informal gathering of small numbers $r$ of people selected randomly. Agents are followers, i.e.,  agents who holding either one of the two stances, are likely to switch stance if given more convincing arguments than those they use to legitimize their current one. This feature is implemented applying a local majority rule within each group of size $r$ to update the individual stances within the group. As a result, the local majority convinces the local minority. Afterward the agents are reshuffled and the precedent process is repeated. 

To keep calculations simple we restrict the meeting groups to a size of three agents ($r=3$). In this case the  update equation writes,
\begin{equation}
p_{t+1}=p_t^3+3(1-p_t)p_t^2 ,
\label{p3}
\end{equation}
which yields a threshold dynamics with a tipping point located at $0.50$. The initial global majority convinces the overall minority via individual shifts \cite{mino}.

In the case of radicalism, R and C are also followers with each agent holding either one of the two stances, extremist or moderate. However, while in the basic model updates are performed systematically within each group at every update, in the case of a sensitive issue agents can decide locally either to engage or to avoid confronting their respective views about the issue. 

Indeed, a topic connected to an on going radicalism produces an asymmetry between C and R with respect to their respective propensity to engage a discussion. C perceived the issue as controversial and consider it is none of their business. They tend to avoid debating the issue with R. At the opposite, R aim at convincing  C foreseeing the spread of their vision as a collective agenda. Aware of their follower status, R apply a strategy in engaging a debate with C in order to maximize their chance of turning some C into R. Accordingly, they force a debate when being locally majority in a discussing group and refrain when in minority.

To account for this asymmetry we denote by $m$ the probability for a group to engage in a local discussion if R is majority and $n$ if C is majority. Given some initial condition $p_0$ at $t=0$ with a set of fixed values $(m, n)$, the opinion dynamics is implemented via repeated cycles of local updates according to the following steps:
\begin{enumerate}
\item All agents are distributed randomly in small groups of fixed size $3$.
\item Within each group a local majority update is made with probability $m$ in case of two R and one C
and probability $n$ for one R and two C. Otherwise no update is performed, i.e., the issue was not discussed.
\item After the local updates, agents are dispersed and reshuffled.
\item At time $t$, $t$ cycles of updates are performed yielding the series,
\begin{equation}
p_0 \rightarrow p_1 \rightarrow p_2 \rightarrow \dots \rightarrow p_t .
\label{dyna1}
\end{equation}
\end{enumerate}

The four steps turn Eq.(\ref{p3}) into the WGM update equation,
\begin{equation}
p_{t+1}=p_t^3+(m+2)(1-p_t)p_t^2+(1-n)p_t(1-p_t)^2 ,
\label{dyna2}
\end{equation}
which allows to build the landscape of the dynamics identifying the associated attractors and tipping points. 

\subsection{Tipping points and dynamics}

The associated fixed point equation $p_{t+1}=p_t$ yields two attractors $p_R=1$ (all agents are R) and $p_C=0$ (all agents are C), separated by a tipping point,
\begin{equation}
p_F=\frac{n}{m+n} .
\label{pf}
\end{equation}
When $p_0<p_F \Rightarrow p_{t_{<F}}\approx 0$ and $p_0>p_F \Rightarrow p_{t_{>F}}\approx 1$ where $t_{<F}$ and $t_{>F}$ are  the number of  local updates required to reach an attractor starting from an initial support $p_0$ respectively below and above the tipping point.

The values for $t_{<F}$ and $t_{>F}$ are obtained as approximate formulas derived from an extension of the corresponding formula obtained in the case of a symmetrical tipping point \cite{corre2} with,
\begin{equation}
t_{<F}\approx \bigg\lceil \frac{1}{\ln \lambda} \ln \frac{p_F}{p_F-p_0} \bigg\rceil +1 ,
\label{n<f}
\end{equation}
and
\begin{equation}
t_{>F}\approx \bigg\lceil \frac{1}{\ln \lambda} \ln \frac{1-p_F}{p_0-p_F} \bigg\rceil +1 ,
\label{n>f}
\end{equation}
where $\lceil x \rceil$ is the ceiling function and $\lambda$ is given by,

\begin{equation}
\lambda \equiv  \frac{\partial p_{t+1}}{\partial p_t} \bigg|_{p_F} .
\label{lambda}
\end{equation}
The larger $t_{<F}$ and $t_{>F}$ the longer is the time required to complete the shrinking or spreading of R.

\begin{figure}
\includegraphics[width=.4\textwidth]{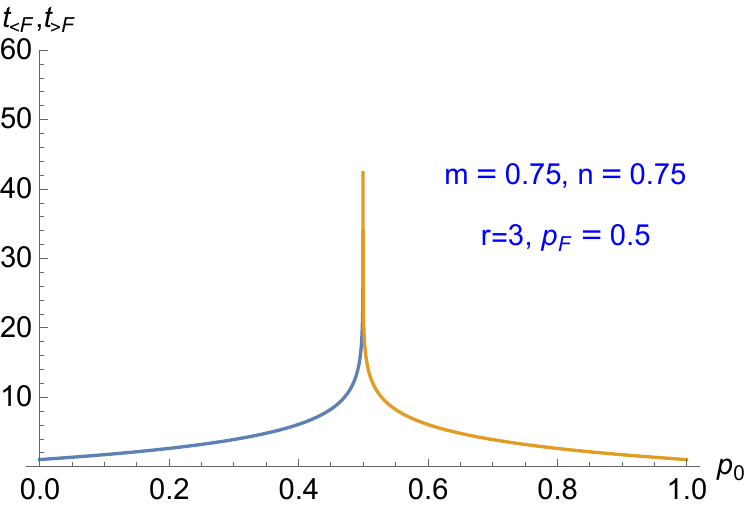}
\includegraphics[width=.4\textwidth]{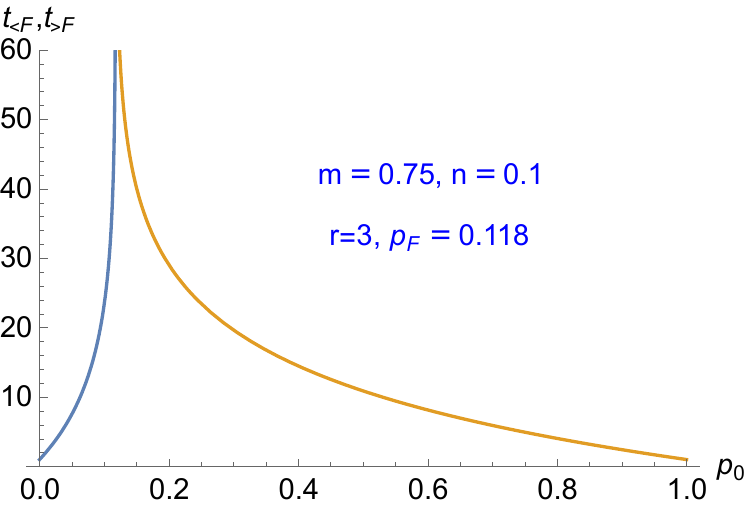}
\includegraphics[width=.4\textwidth]{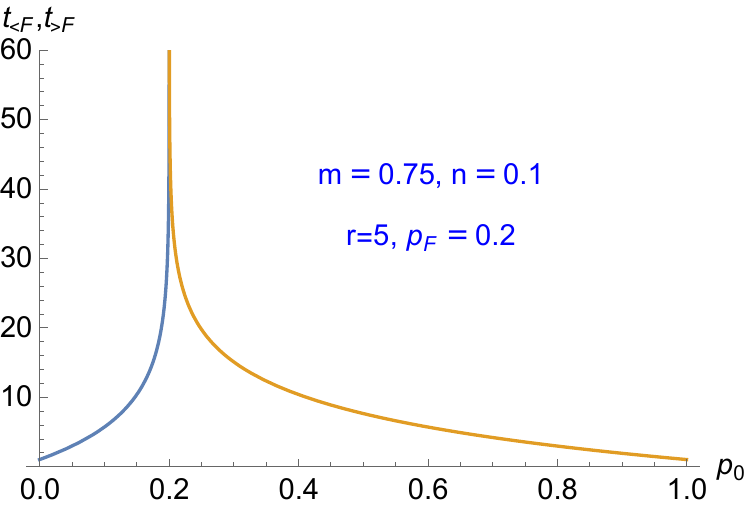}
\caption{Number of updates $t_{<F}$ and $t_{>F}$ using Eqs. (\ref{n<f}, \ref{n>f}) for both $r=3$ and $r=5$ as a function of $p_0$ for a given set $(m, n)$. Upper part has $m= n=0.75$ with $r=3$. Middle part  has $m=0.75, n=0.10$ with $r=3$. Lower part $m=0.75, n=0.10$ with $r=5$.}
\label{t123}
\end{figure}

Figure  (\ref{t123}) shows the variation of both Eqs. (\ref{n<f}, \ref{n>f}) with $r=3$ (upper and middle parts) as a function of $p_0$ 
for the symmetric case $m=n=0.75$ (upper part) and the asymmetric case $m=0.75$ and $n=0.10$ (middle part).

\begin{figure}
\includegraphics[width=.35\textwidth]{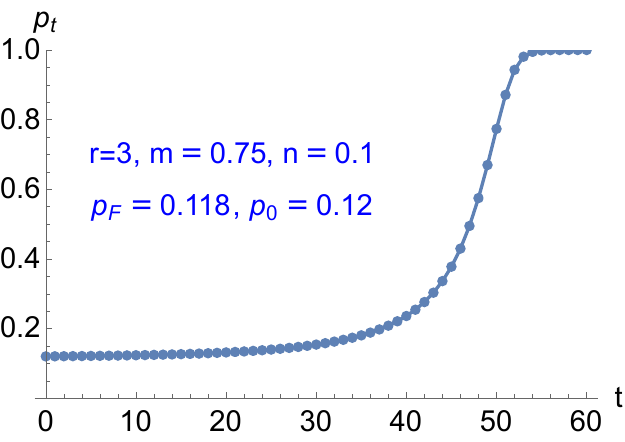}
\includegraphics[width=.35\textwidth]{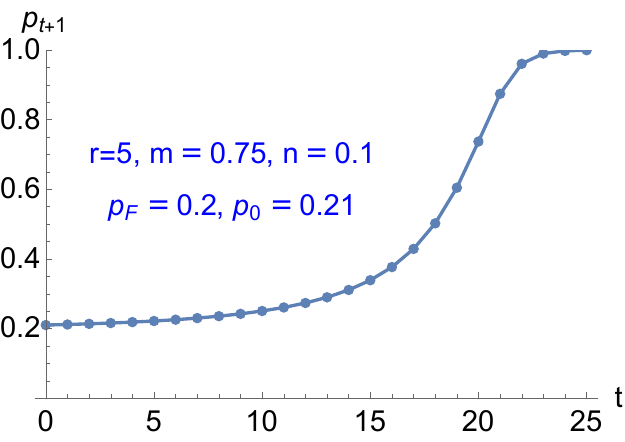}
\caption{Evolution of an initial $p_0>p_F$ with $m=0.75$ and $n=0.10$ as a function of repeated updates with a fixed $r$. Upper part has $r=3, p_0=0.12, p_F=0.118$. Lower part has $r=5, p_0=0.21, p_F=0.20$.}
\label{d12}
\end{figure}

It is seen that the asymmetric case requires many more updates to reach an attractor than the symmetric case. It implies that during long periods of time, i.e. a large number of updates,  the spreading of R is not significant making the ongoing process difficult to notice. After this long period of tiny changes there is a sudden surge in spreading. This sudden change is illustrated in the upper part of Figure (\ref{d12}) which exhibits the evolution of an initial low value $p_0=0.12>p_F=0.118$ of R with $m=0.75$ and $n=0.10$. It is seen that during the first $40$ updates the proportion of R does not increased much. The same feature is observed in the upper part of Figure (\ref{z12}), which shows the variation of $p_{t+1}$ as a function of $p_t$. At low values of $p_t$ up to about $0.30$ the increase of $p_{t+1}$ measured with the distance from the diagonal is very small.

\subsection{Increasing the group size}

Above results are obtained restricting the size of discussing groups to $3$ people. Although this restriction allows to solve the equations analytically it may appear rather narrow. However that is not the case since we are dealing with informal private discussions, which always happen in small groups of people between $2$ and $5$ or $6$. Larger groups fragment spontaneously in smaller subgroups as observed for instance during dinners. Yet to extend our investigation we perform some calculations for larger groups with $r=5$. In this case the update equation Eq. (\ref{dyna2}) becomes,
\begin{equation}
\begin{split}
 \begin{aligned}
p_{t+1} & =  p_t^5 + (4 + m) p_t^4 (1 - p_t) + (6 + 4 m) p_t^3 (1 - p_t)^2 \\
&+ 4 (1 - n) p_t^2 (1 - p_t)^3 + (1 - n) p_t (1 - p_t)^4 ,
 \end{aligned}
 \end{split}
\label{d5-1}
\end{equation}
whose associated numbers of updates to reach an attractor  are shown in the lower part of Figure (\ref{t123}) for $m=0.75$ and $n=0.10$. 

Comparing with $r=3$ (middle part) we found an increase of the tipping point from $p_F=0.118$ to $p_F=0.20$ accompanied with a decrease in the number of updates to reach an attractor, which nevertheless stays higher than in the symmetric case (upper part).

\begin{figure}
\includegraphics[width=.35\textwidth]{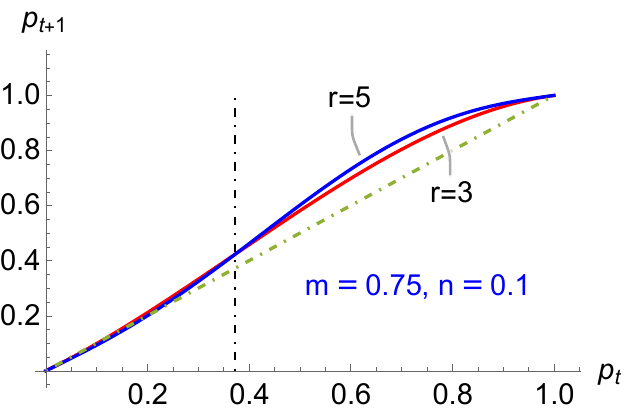}
\includegraphics[width=.35\textwidth]{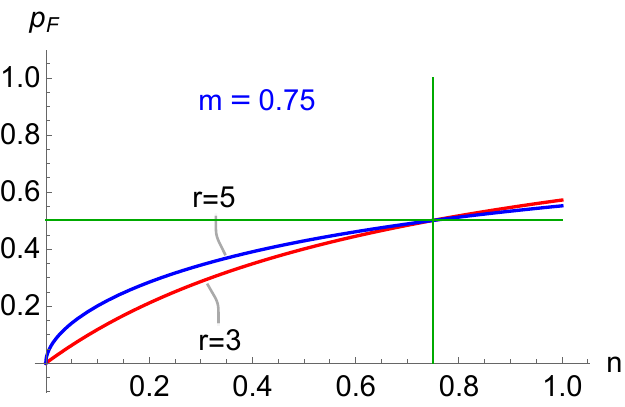}
\caption{
Given $m=0.75$ for both $r=3$ and $r=5$, the upper part shows the functions $p_{t+1}$ as a function of $p_t$ for $n=0.10$ and the lower part exhibits $p_F$ as a function of $n$.}
\label{z12}
\end{figure}

This decrease is also observed in the lower part of Figure (\ref{d12}) with respect to the upper part ($r=3$), The variation of an initial proportion $p_0=0.20$ of R as a function of successive updates, still for $m=0.75$ and $n=0.10$, also shows a sudden surge but occurring about $15$ updates instead of $40$ for $r=3$. Therefore, the qualitative behaviors are identical despite some quantitative changes as equally seen in the upper part of Figure (\ref{z12}), which exhibit $p_{t+1}$ as a function of $p_t$ shows no significant difference between $r=3$ and $r=5$.

To complete our comparison between $r=3$ and $r=5$ we have plotted the value of their respective tipping points as a function of $n$ for a fixed value $m=0.75$ in the lower part of Figure (\ref{z12}). It is seen that $p_F$ is slightly larger for $r=5$ in the range $m>n$. The two values cross at $0.50$ at $m=n$ and then it is $r=3$ which yields a slightly higher value when $m<n$.

Above results show that the main features of the dynamics obtained with groups of three people are qualitatively preserved when going to five people.

\subsection{The phase diagrams}

To uncover the topology of the landscape which drives the dynamics generated by Eq.(\ref{dyna2}), it is fruitful to notice  that the symmetrical case $m=n=1$ locates the tipping point $p_F$ at the symmetrical value $p_F=\frac{1}{2}$. A perfect balanced competition is thus obtained with the initial majority spreading over the minority, either R or C. In parallel, the total asymmetric cases ($m=1,n=0$) and ($m=0,n=1$) lead to every agent respectively R and C independently of $p_0$.

If symmetry was prevailing in real situations, radicalization would rarely be successfully spreading since most cases have initial C large majorities with small R minorities. However, most cases have a broken symmetry with $m\neq n$ making $p_F\neq\frac{1}{2}$. Then, one of the two competing views has an advantage since it can win the competition starting from a minority initial support while the other view can lose starting from a majority initial support. More precisely $m>n \Rightarrow p_F < \frac{1}{2}$ and $m<n \Rightarrow p_F > \frac{1}{2}$. The fate of the dynamics is then sealed by the position of $p_0$ with respect to $p_F$. A larger propensity to engage in local discussions decreases the value of $p_F$, which can compensate a low $p_0$. In contrast, a large $p_0$ allows less involvement in discussing the issue at stake.

Asymmetries between R and C are instrumental in making R likely to spread even starting from very low initial support. In particular:
\begin{description}
\item[(i)] R have a collective agenda to spread their view aiming to convince a maximum of C. Aware of being followers,  they implement their goal applying a selective choice in when to force a debate with C. 

\item[(ii)] C are not interested in debating the issue and have no agenda. They tend to avoid the debate adopting a neutral indifference. C have a low propensity to initiate a debate about the issue even when they could  convince R.  Most C consider that radicalization is none of their business. It is up to the state, public and private spheres to address the phenomenon.
\end{description}

Above asymmetry implies that $n$ is given with C not being aware of its value. On the contrary, R can act to tune $m$ according to both their agenda and initial support. We thus investigate the conditions under which an initial proportion $p_0$ of R spreads against C. 

The making of the value $p_0$ results from opposing external factors, some fueling radicalization and others countering it. Determining the actual value $p_0$ is out of the scope of the present study.  

Alike, given the overall neutral indifference of C combined with the absence of a collective concern, $n$ stems from the distribution of individual propensities within C. It is thus given for each community and is not subject to change at short scales of time. In contrast, R can modify $m$ at their will. It is a decisive advantage in the associated unfolding of radicalism. Afterwards, knowing the conditions leading to the successful spreading of R, allows in turn to build paths to thwart the spreading. 

Therefore, given $p_0$ the parameters $m$ and $n$ set the fate of the dynamics for R either to spread or shrink. Their values are a function of the respective propensities of R and C to engage or to avoid discussing the underlying issue when gathered during informal private meetings. 

\section{The spread of Radicals}

\begin{figure}
\includegraphics[width=.4\textwidth]{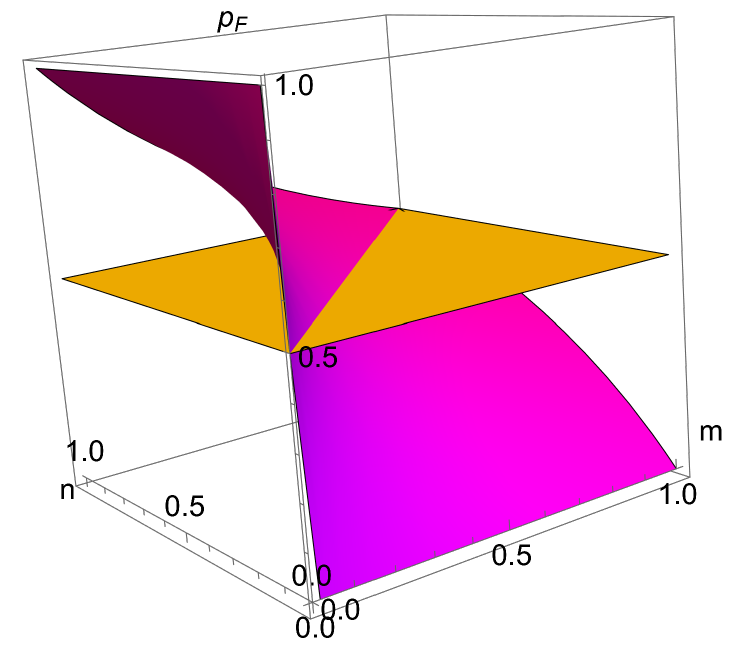}
\caption{\label{pf1} The tipping surface $p_F=\frac{n}{m+n} $ as a function of $m$ and $n$. When $p_0$ is above the surface, radicalization spreads and it shrinks below. In the area between the tipping surface (above) and the plane at $\frac{1}{2}$ (below) radicalization spreads starting from a minority support.} 
\end{figure}

Figure (\ref{pf1}) shows the variation of $p_F$ as a function of $m$ and $n$ using Eq. (\ref{pf}). When $p_0 >p_F$ R spread and shrink for $p_0<p_F$.

However, the values $n$ and $p_0$ being fixed over some period of time, it is more appropriate to recast $p_0>p_F$ as,
\begin{equation}
m_c=n \frac{ 1-p_0}{p_0} ,
\label{mc}
\end{equation}
which yields R spreading for $m>m_c$.

Eq. (\ref{mc}) indicates that when $m_c>1$, the condition $m>m_c$ cannot be satisfied, which means that R cannot spread and thus shrinks. R spreading is thus impossible when either,
\begin{equation}
n >n_c\equiv \frac{ p_0}{1-p_0} ,
\label{m1}
\end{equation}
or
\begin{equation}
p_0<p_c\equiv\frac{n}{1+n} ,
\label{m2}
\end{equation}
where $p_c$ is the value of $p_F$ at $m=1$.

The first inequality corresponds to the viewpoint of C yielding its minimum value of involvement $n$ to block the spreading of R given an initial support $p_0$. The second inequality sets the minimum value $p_0$ below which R spreading is impossible with a certain shrinking.

\begin{figure}
\includegraphics[width=.4\textwidth]{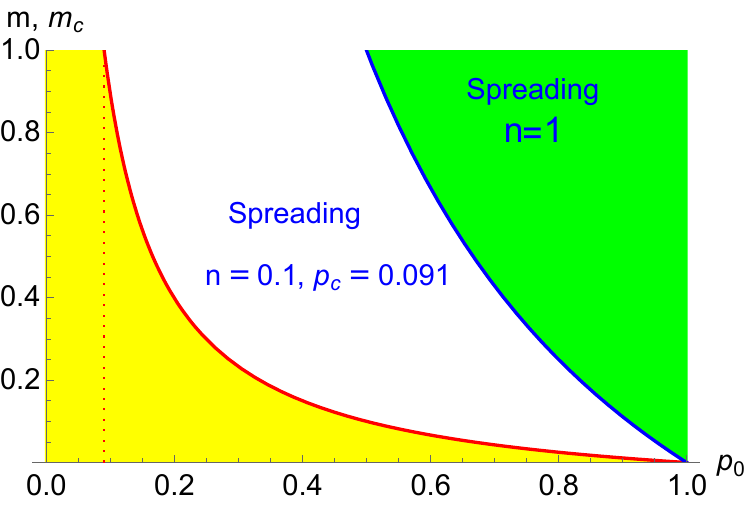}
\includegraphics[width=.4\textwidth]{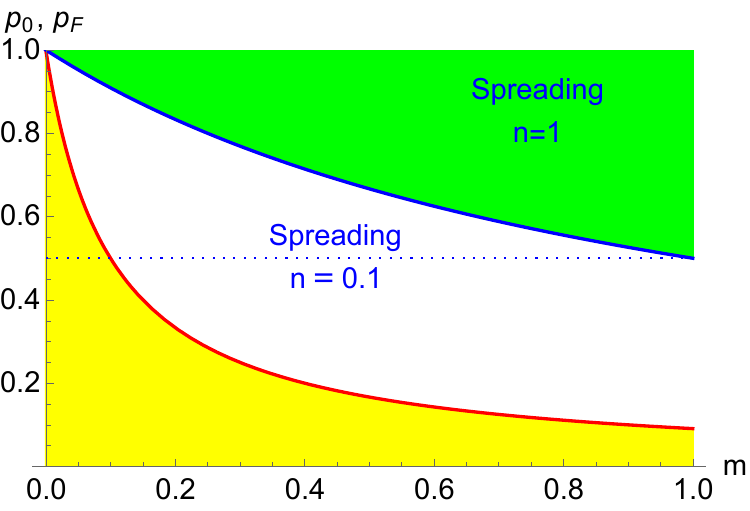}
\caption{Landscape of R spreading at $n=0.10$ and $n=1$. The upper part shows both frontiers $m_c=n\frac{1-p_0}{p_0}$ as a function of $p_0$. Beyond the solid curve R spread. The case $n=0.1$ includes the white and dark areas while it is only the dark area for $n=1$. The lower part shows the same spreading areas but from the perspective of the frontier $p_F=\frac{n}{n+m}$ as a function of $m$.}
\label{v12}
\end{figure}

The extreme case $n=1$ illustrates the harder conditions for R spreading with Eq. (\ref{m2}) giving the inequality $p_0<\frac{1}{2}$. Without an initial majority R is set to shrink. However the condition $p_0>\frac{1}{2}$ does not guarantee R spreading, which requires the additional condition $m> m_c$. With $n=1$, it yields $m>\frac{ 1-p_0}{p_0}$ as shown in the upper part of Figure (\ref{v12}). The lower part of the Figure shows the  same landscape with $p_0$ as a function of $m$ instead of $m$ as a function of $p_0$.

Having determined the $n=1$ upper limit landscape it is worth to explore the variation of $p_F$ as a function of $m$ for different values of $n$ since $p_F$ is the minimum support required to have the possibility of R spreading. It is equivalent to have $m_c$ as a function of  $p_0$ given $n$. Figures  (\ref{v12}) exhibit those landscapes for $n=0.10$.

\begin{figure}
\includegraphics[width=.4\textwidth]{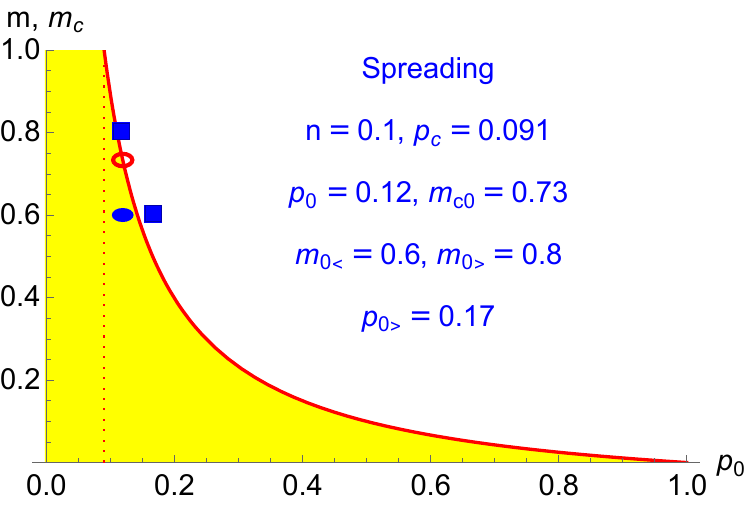}
\includegraphics[width=.4\textwidth]{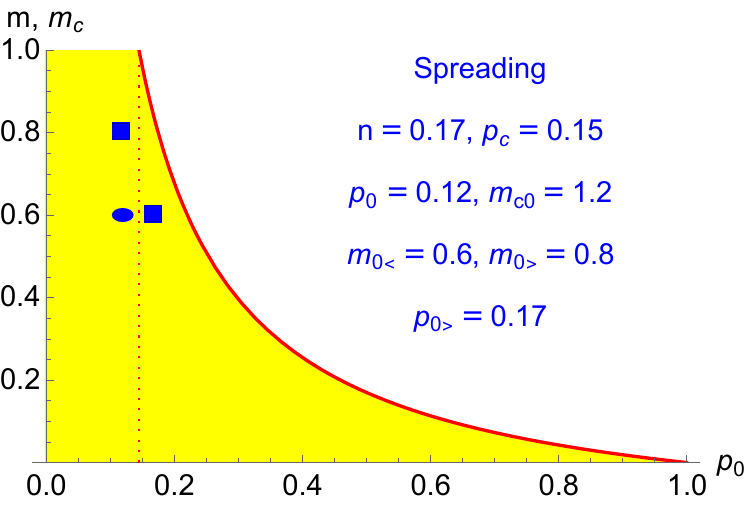}
\caption{
Landscape for radicalization showing the frontier $m_c=n\frac{1-p_0}{p_0}$ between the shrinking (dark area) and spreading (white area) zones as a function of $p_0$. The vertical dotted line delimits on its left side the impossible spreading sub-zone when $p_0<p_c$.The upper part has $n=0.10$ with $p_c=0.091$. Three points are included with $p_0=0.12$ and respectively $m_{0<}=0.60$ (full circle, shrinking), $m_{c0}=0.73$ (empty circle, frontier), $m_{0>}=0.80$ (full square, spreading). A fourth point is included at $p_{0>}=0.17$ and $m_{0<}=0.60$ (full square, spreading).The lower part is similar to the upper part with $n=0.17$, which expands the shrinking zone with $p_c=0.15$. The two points with $p_0=0.12<p_c$ are now in the impossible spreading sub-zone and the point with $p_{0>}=0.17$ in the shrinking zone.
}
\label{v34}
\end{figure}

Indeed, the faith of R depends on the interplay between $m$ and $(n, p_0)$, which sets the dynamics towards either spreading ($m>m_c$) or shrinking  ($m<m_c$). On this basis, winning strategy for R can be identified starting from initial $p_0$ and $m$ for a given $n$. 

The upper part of Figure (\ref{v34}) illustrates a situation with $n=0.10$, which yields $p_c=0.091$. Accordingly, any R initial support below $p_c$ is doomed to shrink even with the highest propensity $m=1$. However, selecting a $p_0>p_c$ is not sufficient as seen in the Figure (upper part) with the point $p_0=0.12$ and $m_{0<}=0.60$. To be located in the spreading zone requires the additional condition $m>m_c$ to be satisfied. In our case ($p_0=0.12, m_{0<}=0.60$) two paths are available for R reaching the spreading zone.

The first path is for R to focus on increasing the propensity to discuss the issue at the value $m > m_c=0.73$ to locate the initial support in the spreading zone,  like for instance with $m_{0>}=0.80$. The second path is to keep unchanged the propensity $m_{0<}=0.60$ but to increase the initial support from $p_0=0.12$ to a higher value located in the spreading zone like with $p_{0>}=0.17$.

Above examples illustrate the easiness for R to spread over, despite having a low initial support $p_0$ taking advantage of the substantial asymmetry in their propensity to engage in local debate about their collective goal in contrast to C inclination to avoid confronting the controversial issue.

\section{The Conventional key to counter Radical spreading}

Figure (\ref{v12}) and the cases exhibited in the upper part Figure (\ref{v34}) have allowed to identify the two key strategies for R to spread within an overwhelming majority C without the need to apply any coercion, neither external nor internal.  

Both stem from the passive indifference of C with respect to the issue raised by R. A low value of $n$ produces low barriers of initial R support and required propensity to debate for R to spread. Depending on $n$, it is sufficient for R to adjust either $p_0$ or and $m$ appropriately.

In contrast, those instrumental advantages of R shed light also on operative paths for C to counter R spreading. In particular, it is found that a stronger C involvement in confronting back R in local gathering is an absolute prerequisite to any efficient policy to curb R spreading. As seen from our results, to rely only on official institutions and the state to reduce the value of $p_0$ is not enough, especially since it is impossible to reduce R support to very low values. Therefore increasing $n$ is an essential necessity, which is however not too much demanding.

To sustain above statement the lower part of Figure (\ref{v34}) shows the modified spreading landscape with $n=0.17$ instead of $n=0.10$. A slight $0.07$ increase in C propensity to debate puts the precedent  spreading locations ($p_0=0.12, m_{0>}=0.80$) and ($p_{0>}=0.17, m_{0<}=0.60$) into the shrinking zone. The slight $n$ increase makes $p_c$ to go from $0.09$ up to $0.15$. In addition $p_0=0.12$ gives $m_c=0.83$.

Indeed, the impossible spreading zone increases substantially with increasing $n$. For instance, while $n=0.10$ gives $p_c=0.09$, $n=0.30$ yields $p_c=0.23$ and $n=0.60$ gives $p_c=0.38$ putting the minimum initial value for R spreading at a high value, which is difficult to reach requiring a very strong external pressure to get large proportions of C to shift lonely to R without interacting on the issue. Figure (\ref{v12}) shows the drastic reduction of the spreading zone going from $n=0.10$  (white area plus upper dark area) to $n=1$ (only dark area).

\section{Conclusion}

We have studied the effect of an asymmetry in the individual propensities to engage in a social debate to put at stake a moderate practice to replace it with an extreme version within a community.  The extremist option is advocated by Radicals (R) to convince Conventionals (C) who are initially the large majority in the community. Using a Weighted Galam Model of opinion dynamics (WGM) has shed light on the ease with which R spread even when starting with a rather low initial support in the community.

The results  provide a rationale to three salient puzzling questions that characterize the phenomenon of radicalization, which are:
\begin{description}
\item[(i)] Why a handful of Radicals is sufficient to start a process, which will eventually reach a substantial part of a community
whose members were mostly following an established conventional practice?
\item[(ii)] Why the spreading goes smoothly and unnoticed during long periods of time before appearing at once visible and massive?
\item[(iii)] Why the phenomenon is resilient against repressive measures coming either from public or private spheres? 
\end{description}

These findings indicate that only a change in attitude from C can put a stop to the dynamics of radicalization. C must accept to tackle the issue as being their issue by engaging against R when possible in private informal meetings. They must adopt a collective vision to the problem in a similar manner R do. Moreover, they must become aware of their a priori important advantage. Being the large majority in their community their individual involvement can be much less demanding than what is required from Radicals when starting as a tiny minority.

Last but not least, it is of importance to underline that not every extremist vision will automatically spread. All our results are based on the instrumental assumption that a local majority of R do convince a minority of C provided they accept to discuss the issue raised by R. This assumption holds only when R do have convincing arguments and/or when C are sensitive to the arguments put forward in the informal discussions. The effective values of the probabilities $m$ and $n$ do account for this convincing power by combining the propensity to engage in a discussion and the strength of the arguments.

In a future work we intend to apply the WGM to investigate the dynamics of Evangelists in their campaign to convert Conventionals.

\section*{Acknowledgment} We thank Gabor T\`oth for a careful reading of the manuscript.


\end{document}